\documentclass[]{aa}
\usepackage{natbib}
\bibpunct{(}{)}{;}{a}{}{,}
\usepackage{graphicx}
\usepackage{subfigure}
\usepackage{txfonts}

\begin{document}

\title{Spectroscopic monitoring of the transition phase in 
nova V4745~Sgr\thanks{Based
on data obtained at the Mount Stromlo and Siding Spring Observatories, 
Australia and the Wise Observatory, Israel}}

\author{B. Cs\'ak\inst{1}
	\and
	L.~L. Kiss\inst{2}\thanks{On leave from University of Szeged, Hungary}
	\and
	A. Retter\inst{3}
	\and
	A. Jacob\inst{2}
	\and
	S. Kaspi\inst{4}}

\institute{Department of Experimental Physics and Astronomical Observatory,
University of Szeged,
Szeged, D\'om t\'er 9, 6720 Hungary
\and
School of Physics, University of Sydney 2006, Australia
\and
Department of Astronomy and Astrophysics, Pennsylvania State University,
525 Davey Laboratory, University Park, PA 16802-6305, USA
\and
School of Physics and Astronomy and the Wise Observatory, 
Tel Aviv University, Tel Aviv 69978, Israel
}

\titlerunning{Spectroscopy of the transient nova V4745~Sgr}
\authorrunning{B. Cs\'ak et al.}
\offprints{B. Cs\'ak,\\
 e-mail: {\tt csakb@physx.u-szeged.hu}}
\date{}

\abstract{We present a spectroscopic monitoring of the transient nova
V4745~Sagittarii (Nova Sgr 2003~\#1) covering ten months after the
discovery. During this period the light curve showed well expressed
transient phase in the form of six rebrightenings, and the presented 
dataset is one of the best spectroscopic coverages of a classical nova
during the transition phase.  Low- and medium-resolution spectra 
revealed that the majority of  spectral lines switched back to strong
P-Cyg profiles during the  mini-outbursts, similar to the spectra just
after the major eruption. We interpret the observed phenomena as
evidence for ``echo-outbursts''  resulting in episodes of secondary
mass-ejections and propose that the  transition phase in classical
novae is driven by repetitive instabilities of the hydrogen shell
burning on the surface of the white dwarf.  From the available light
curve data we  estimate the absolute magnitude of the system of about
$-8\fm3\pm0\fm5$.  All spectrophotometric pieces of evidence point
toward a very low  ($E(B-V)<0\fm1$) interstellar reddening, leading to
a rough distance  estimate of V4745~Sgr ($9~{\rm kpc}<d<19~{\rm
kpc}$). 

}

\maketitle

\section{Introduction} 

V4745~Sgr (Nova Sgr 2003~\#1; $\mathrm{R.A. =18^{h}40^{m}02\fs54}$,
$\mathrm{Decl. = -33\degr26\arcmin55\farcs1}$, equinox J2000.0) was
discovered independently by Brown and Yamamoto \citeyearpar{iauc8123}
on T-Max 400 films on Apr. 25 and 26 at $m_{\rm pg}=9\fm6$. Nothing was
visible to 10\fm5 at this location on Yamamoto's patrol films taken
during 2000 Apr. 28--2003 Apr. 5. Spectroscopic confirmation was given
by optical and infrared spectroscopy \citep{ashok03}. The first optical
spectrum showed it was a nova of the ``Fe~II'' class in the
classification system of \cite{williams92}, with strong Balmer lines
having P-Cyg profile. After rapid fading it rebrightened again within a
few days, and the emission lines of the Balmer series and strong Fe II
emission  lines at 4176 and 4233~\AA\ showed complex P-Cyg profiles
\citep{iauc8132k}.

In Fig.~\ref{fig:lc_spec} we show the light curve of the nova based on
visual,  photographic and CCD observations collected by the VSNET group
\citep{vsnet}, which includes several pre-discovery observations, too.
From this curve we estimate that the visual maximum occurred at 7\fm4
on 2003 Apr. 18.4 UT. The uncertainty of the maximum brightness is at
least 0\fm5 and 0\fd5 in time, due to the poor observational coverage.
The shape of the light curve -- the maximum brightness, a rapid
decline, followed by rebrightenings which are superimposed on the
general fading -- is very similar to nova V443 Sct
\citep[][]{anupama92}, and other novae with transient type light
curves. 

There are two major types of transient effects in the classical nova 
light curves during decline: the light curve can either show a deep
minimum (like in DQ~Her) interpreted by the formation of an optically 
thick dust envelope around the binary system, or slow, oscillation-like
features of ambiguous origin (e.g., V603~Aql, V1494~Aql, DK~Lac,
GK~Per, V373 Sct, V443~Sct, etc). For simplicity, throughout the paper
we  only refer to the latter as the transition phase. 

Several models for the oscillations during the transition phase have
been proposed, e.g.,  steady-state super-Eddington winds
\citep{shaviv01}, oscillations of the common envelope, or the hot white
dwarf \citep[see ][ for a recent summary]{retter02}, but there are
still not enough observational studies to give a firm explanation of
this phenomenon. Very recently, \cite{retter02} proposed a new solution
invoking a possible  connection between the transition phase and
intermediate polars.

The little observational data during the transition phase and the
presence of several competing models motivated this work. Here we
present a spectroscopic monitoring of V4745~Sgr, which resulted  in one
of the best covered transient phases. We detected remarkable spectral
variations, which shed new light on the possible physical mechanism
driving  the light curve ``oscillations''. The main goal of this paper
is to present  our spectroscopic analysis based on observations taken
at and around three  of the rebrightenings between $\Delta$t=26 to 182
d ($\Delta$t refers to the number of days passed since the  optical
maximum). In addition, important physical parameters of the system were
also estimated.

\section{Observations} 


\begin{table}
\caption{The journal of observations. Spectra obtained during a
rebrightening are marked by asterisk.}
\begin{tabular}{lllcccl}
\hline
$\Delta$t &	& date  & range & im. scale & telescope\\
	  &	& 	& $\AA$ & $\AA$/px  &       \\
\hline
26$*$  	  &2003	&May 13 & 3500--4450 & 0.55 & SSO 2.3\\
	  &	&	& 6360--7000 & 0.55 &        \\
28$*$     &---	&May 15 & ---"---    & 0.55 & SSO 2.3\\
31	  &---	&May 18 & ---"---    & 0.55 & SSO 2.3\\
34	  &---	&May 21 & ---"---    & 0.55 & SSO 2.3\\
75$*$ 	  &---	&Jul 1  & 4000--7800 & 3.75 & WO 1.0\\
83	  &---	&Jul 9  & ---"---    & 3.75 & WO 1.0\\
105	  &---	&Jul 30 & ---"---    & 3.75 & WO 1.0\\
111	  &---	&Aug 6  & ---"---    & 3.75 & WO 1.0\\
131	  &---	&Aug 26 & ---"---    & 3.75 & WO 1.0\\
159	  &---	&Sep 23 & ---"---    & 3.75 & WO 1.0\\
173	  &---	&Oct 7  & 4000--4980 & 0.55 & SSO 2.3\\
	  &	&	& 6150--7150 & 0.55 &        \\
174	  &---	&Oct 8  & ---"---    & 0.55 & SSO 2.3\\
175	  &---	&Oct 9  & ---"---    & 0.55 & SSO 2.3\\
182$*$ 	  &---	&Oct 16 & ---"---    & 0.55 & SSO 2.3\\
296	  &2004	&Feb 9  & 5980--7008 & 0.55 & SSO 2.3\\
\hline
\end{tabular}
\label{tab:joo}
\end{table}

\begin{figure} 
\centering 
\includegraphics[width=88mm]{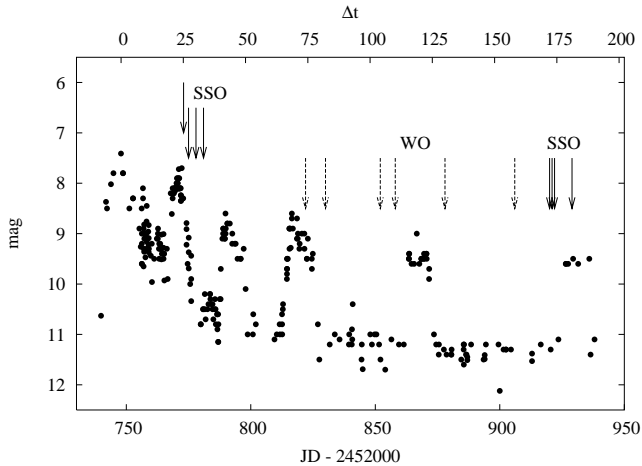}
\caption{The light curve of V4745 Sgr from the VSNET data for the time
interval 2003 April to October. The epochs of spectroscopic observations are
marked with arrows (the last one is not shown).} 
\label{fig:lc_spec}
\end{figure}

The observations were carried out in two locations, on 15 nights
between  May 13, 2003 and Feb. 9, 2004. Medium-resolution spectra (the
image scale was  0.55~$\AA$~px$^{-1}$) were taken with the Double Beam
Spectrograph \citep{dbs88}, mounted in the Nasmyth focus of the 2.3-m
Advanced Technology Telescope  at Siding Spring Observatory (SSO),
Australia. 

Low-resolution spectra (image scale: 3.75~$\AA$~px$^{-1}$) were taken
using the Faint Object Spectrograph and Camera attached to the 1.0-m
telescope of the Wise Observatory (WO), Israel. The observations
details are given in Table~\ref{tab:joo}, while the epochs of
observations are marked in Fig. \ref{fig:lc_spec} with arrows.

All spectra were reduced with standard IRAF\footnote{IRAF is
distributed by the National Optical Astronomy Observatories, which are
operated by the Association of Universities for Research in Astronomy,
Inc., under cooperative agreement with the National Science
Foundation.} tasks, including bias and flat field corrections, aperture
extraction and wavelength calibration. The exposure times varied
between 120 and 900 seconds, depending on the actual wavelength range
and the brightness of emission features.

The SSO spectra were not flux calibrated -- except for two nights in
October -- due to observing conditions, while the WO spectra were
calibrated, using some general flux calibrators observed at the
observatory back to April, so that the flux-level uncertainty can
exceed 10\%. Nevertheless, the overall spectroscopic appearance is
hardly affected by this uncertainty. 

\section{Description of the spectra}

\begin{figure} 
\centering 
\includegraphics[width=88mm]{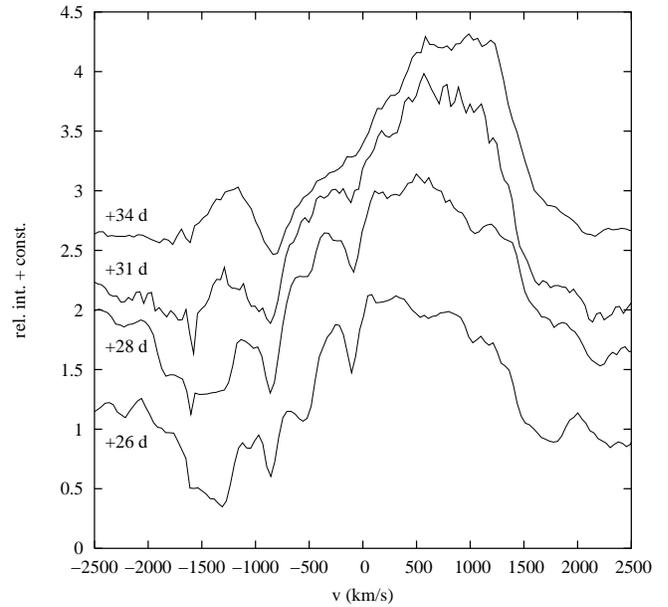}
\caption{The evolution of the H$\gamma$ line during the descending branch of
the second rebrightening.} 
\label{fig:gme}
\end{figure}

\begin{figure*}
\centering
\mbox{\subfigure{\includegraphics[width=88mm]{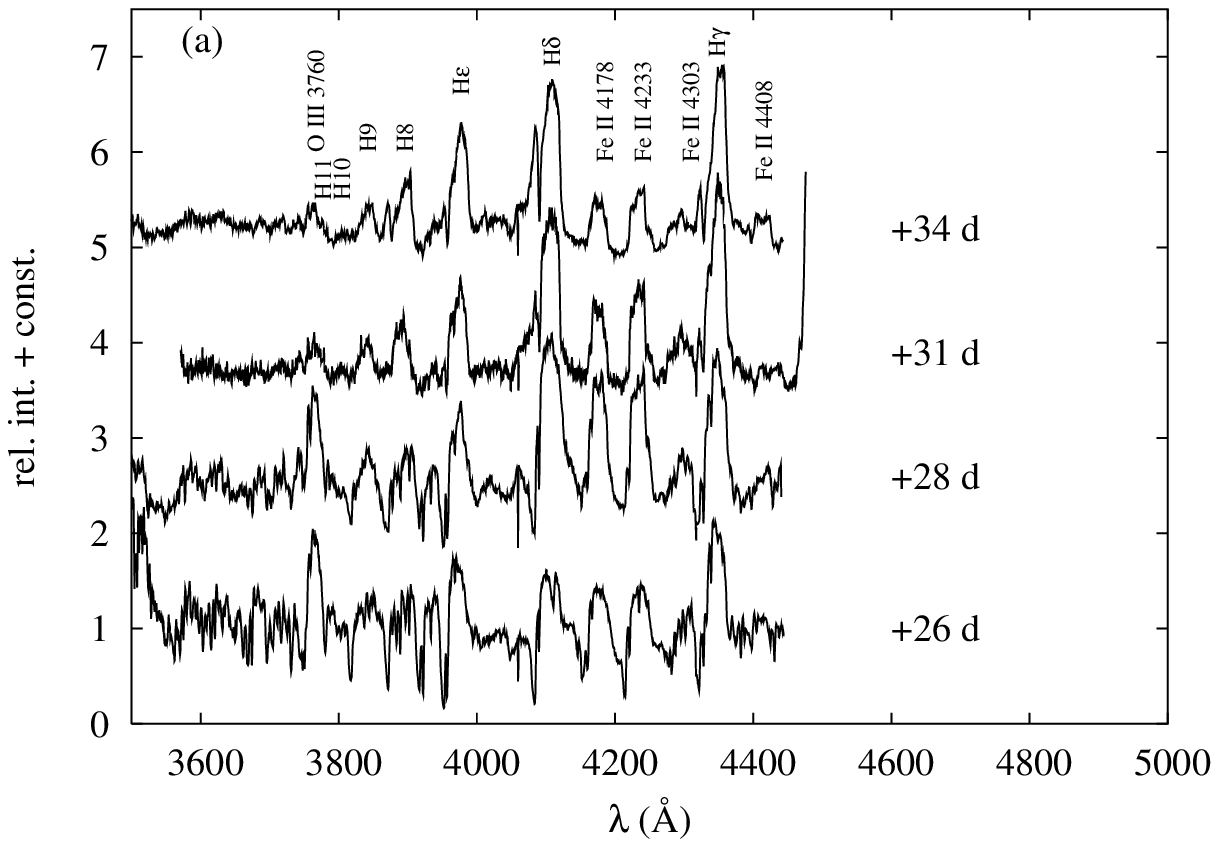}}
	  \subfigure{\includegraphics[width=88mm]{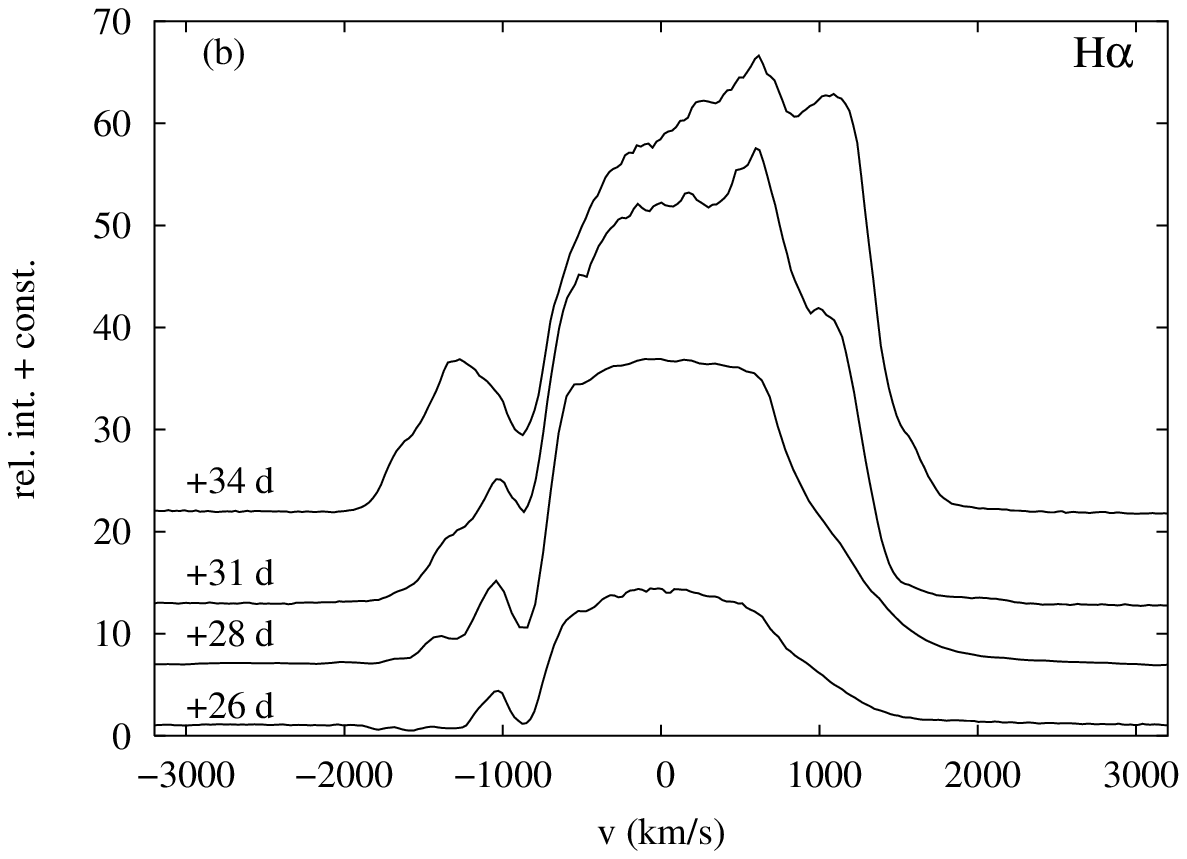}}}
\mbox{\subfigure{\includegraphics[width=88mm]{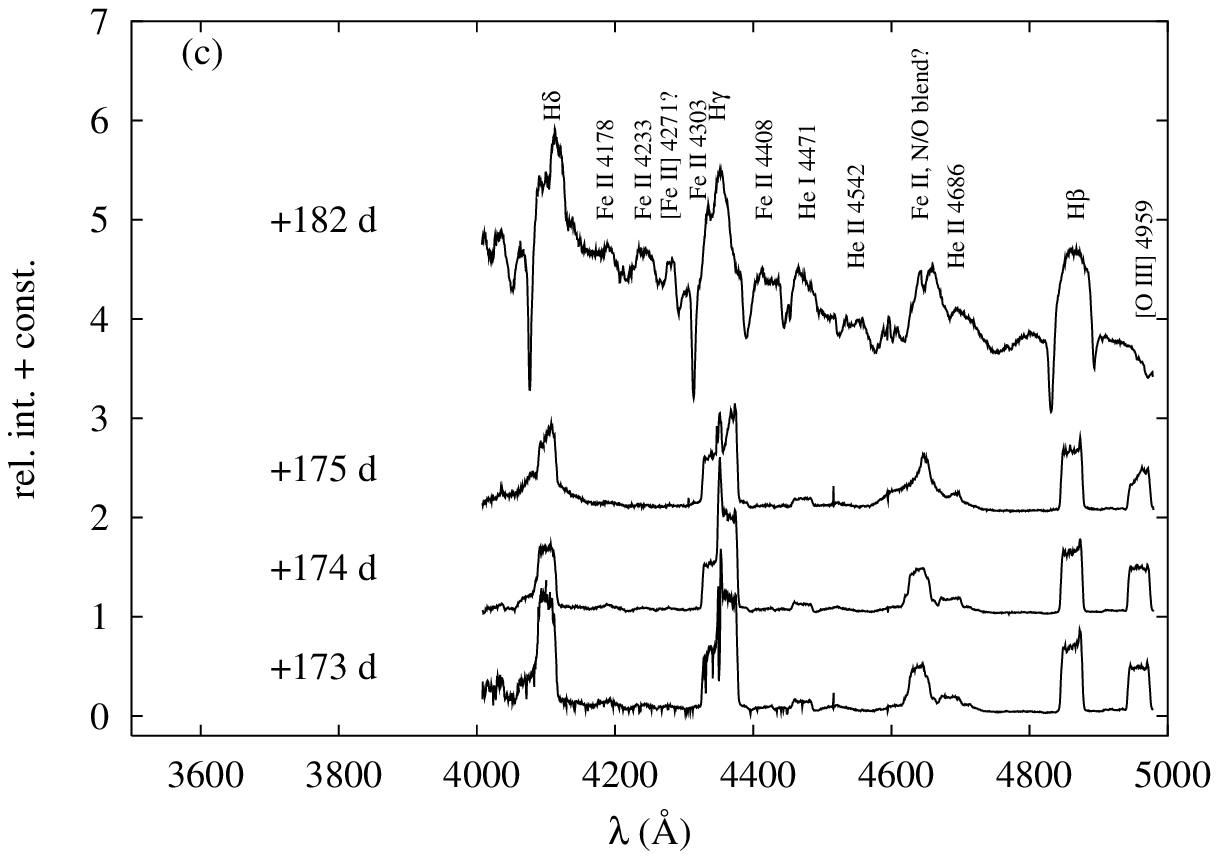}}
	  \subfigure{\includegraphics[width=88mm]{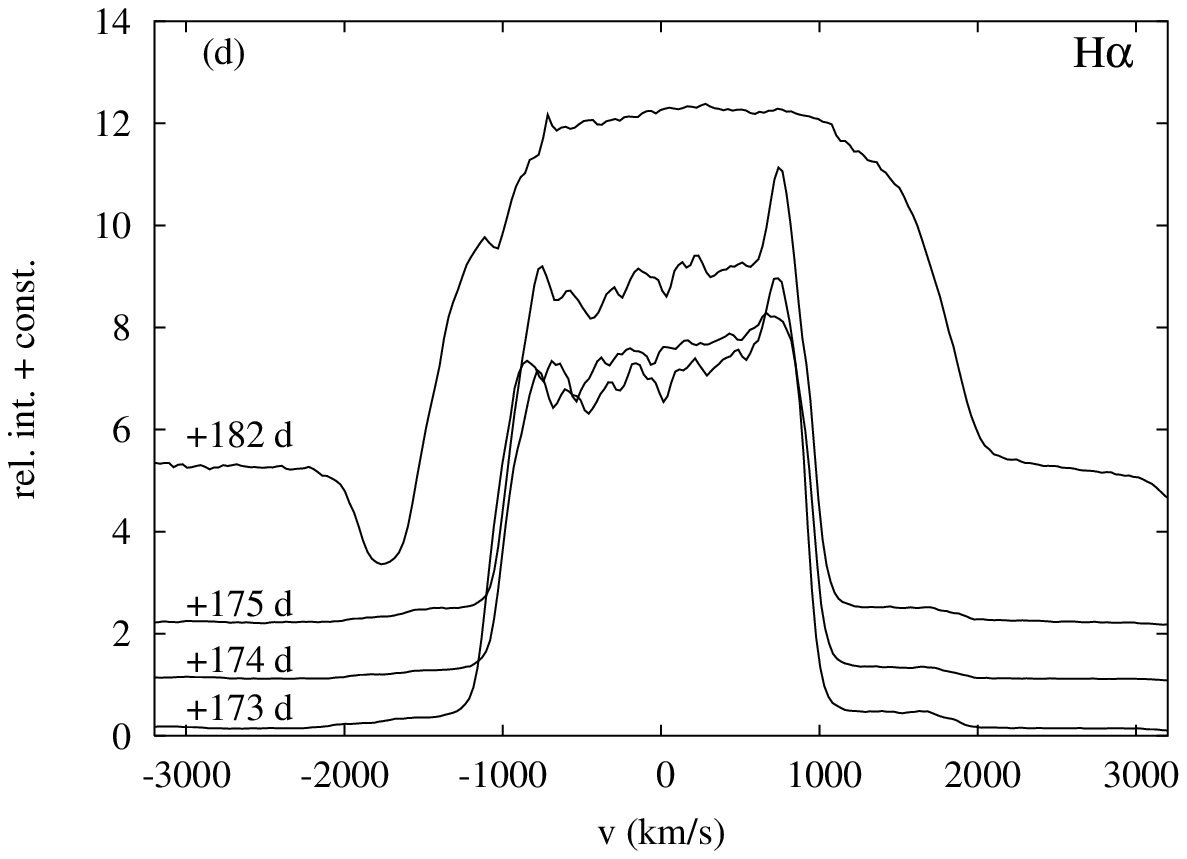}}}
\caption{Two snapshots of the long-term spectral evolution. The two
left panels show the blue spectral region in May and October, 2003, while 
the velocity structure of the H$\alpha$ line is traced in the two right panels.}
\label{fig:specevol}
\end{figure*}

The first set of spectra was obtained in the time interval from 26 to
34 days after the maximum brightness, during the second rebrightening
and the following fading.  The blue spectra, covering 3500~\AA\ to
4450~\AA, are characterized by strong emissions of the  hydrogen Balmer
series and Fe~II lines, all with P-Cyg profiles. The radial velocities
of the absorption features are approximately $-1450$, $-900$, $-500$
and $-150$ km~s$^{-1}$ (the bottom curve in Fig.~\ref{fig:gme}). These
absorption features are also present in the other Balmer and Fe~II
emission lines (see Fig.~\ref{fig:specevol}a). 

\begin{figure*}
\centering
\includegraphics[width=176mm]{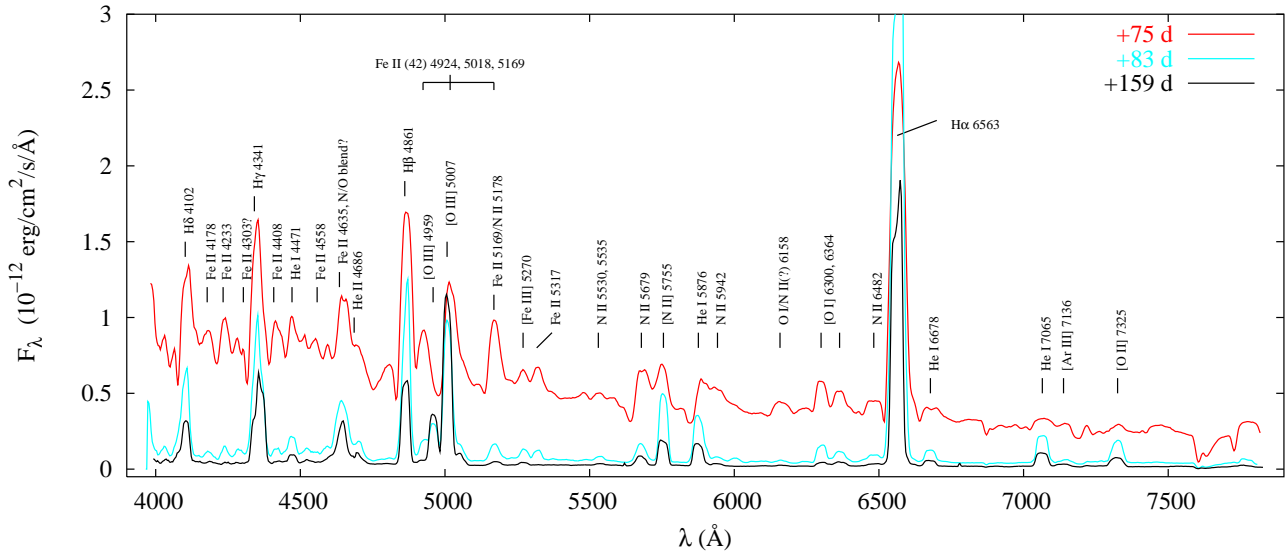}
\caption{Low-resolution spectra of V4745~Sgr 75, 83 and 159 days after
the maximum.}
\label{fig:wise_lines}
\end{figure*}


We obtained spectra simultaneously in the red region, centered
approximately on the H$\alpha$ line. The line profiles are plotted in
Fig.~\ref{fig:specevol}b. Similar absorption features at the same
velocities can be seen in the H$\alpha$ emission line. The relatively
smooth line shape evolved into a broad and complex emission profile. We
can identify strong, redshifted emission features in spectra taken at
$\Delta t = 31$ and 34 days, at velocities $\approx+620$, $+1050$ and
$+1500$ km~s$^{-1}$. These emissions are probably the symmetric pairs
of the blue-shifted absorption profiles. The presence of P-Cyg profiles
suggests that the rebrightening of the nova was accompanied with
mass-ejection. The ejection was not isotropic as implied by the
appearance of complex absorption--emission pairs.

During the following four months we collected low-resolution spectra
at  the WO. A sample of these spectra is plotted in
Fig.~\ref{fig:wise_lines}. We can identify the Balmer series from
H$\alpha$ to H$\delta$, He~I and He~II, forbidden and permitted
O~I/II/III and N~II lines and a couple of Fe~II emissions. Note the
presence of  the nebular lines (forbidden oxygen, nitrogen and iron)
two and half months after the maximum brightness. On July 1 ($\Delta t
= 75~\mathrm{d}$), the nova was six days after its fourth
rebrightening, and we can identify P-Cyg profiles at the blue wings of
the brighter emission lines again at $\approx-1600$ km~s$^{-1}$ (the
upper curve in Fig.~\ref{fig:wise_lines}). In addition, the continuum
is significantly brighter and bluer than in the other phases. At
$\Delta t = 75~\mathrm{d}$, the continuum at the visual photometric
band was 5--6 times brighter than later, at $\Delta$t=83~d. This
behaviour is consistent with the $\sim$2 mag brightening of the system
in the visual band (see Fig.~\ref{fig:lc_spec}).

The next observing run was between Oct. 7 and 16, at the SSO. We
obtained spectra on three consecutive nights, just before the sixth
rebrightening, and on one night that was almost at the maximum
brightness of this minor outburst. The spectra of the blue region are
presented in Fig.~\ref{fig:specevol}c, while the variation of the
H$\alpha$ line is plotted in Fig.~\ref{fig:specevol}d. The changing
emission line profiles (most strikingly in the Fe~II/N/O blend around
4635~$\AA$) and the sudden appearance of broad emission wings at
$\Delta$t=175 d showed that a major change of the system was
developing. Two days after this observation, the sixth rebrightening
was detected by visual observers. The top curves in panels c and d in
Fig.~\ref{fig:specevol} show the outburst features on Oct. 16
($\Delta$t=182 d). The FWHM of the Balmer lines increased from
2000~km~s$^{-1}$ to 3000~km~s$^{-1}$ and strong P-Cyg profiles were
developed. We can identify absorption features of the H$\alpha$ profile
at $-1780$ and $-1030$~km~s$^{-1}$ and a relatively strong [Fe~II]
emission line at 4245~$\AA$. The presence of [Fe~II] lines is very rare
in nova spectra, and we know only several similar cases, for example
CP~Pup \citep{weaver44} and V1494~Aql \citep{iijima03}.

The last observation was made on Feb. 9, 2004 ($\Delta$t=296~d) at the
SSO. The H$\alpha$ line profile is presented in Fig.~\ref{fig:frv}. The
FWHM of the line decreased to approximately 2000~km~s$^{-1}$, and the
emission structure evolved to remarkable triple symmetry. This
structure can be explained with rings of enhanced brightness in the
ejected shell \citep{gill99}. We can easily identify two pairs of
emissions (at --830, 160 and --220, 690~km~s$^{-1}$), which probably
sign the presence of two tropical rings (indicated with straight
lines). We also marked the third, very weak pair with dotted lines
(--580, 460~km~s$^{-1}$). This pair is likely indicate an equatorial
ring.

\begin{figure}
\centering
\includegraphics[width=88mm]{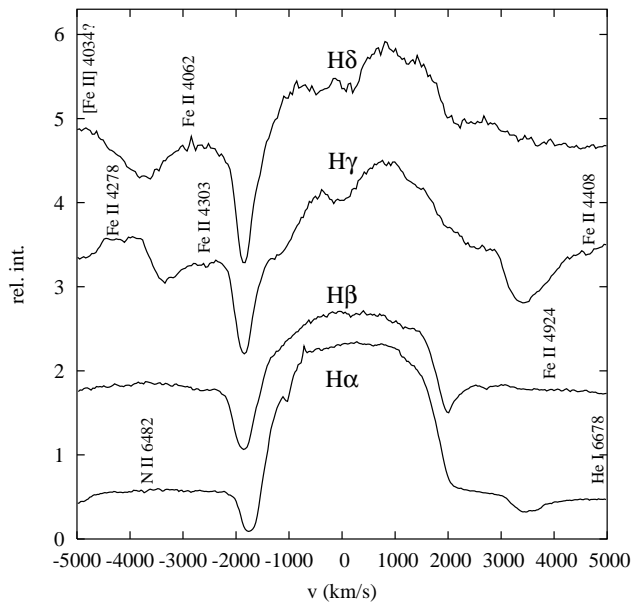}
\caption{The velocity profiles of Balmer lines on 16 Oct. 2003, close to the 
maximum brightness of the sixth rebrightening.}
\label{fig:jetfig}
\end{figure}

\begin{figure}
\centering
\includegraphics[width=88mm]{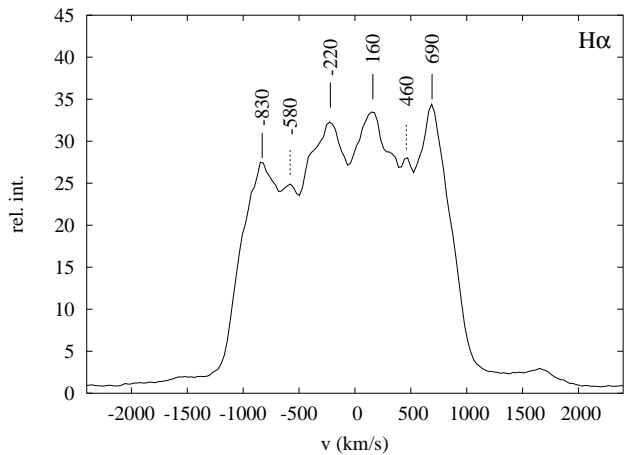}
\caption{The velocity profile of the H$\alpha$ line on Feb. 9 2004.}
\label{fig:frv}
\end{figure}

\section{Physical parameters and the light curve}

\subsection{Absolute magnitude in maximum}

The light curve of V4745~Sgr (Fig.~\ref{fig:lc_spec}) indicates that
the nova reached maximum brightness on April 18.4 UT (JD $2452747.9 \pm
0\fd5$) with $\mathrm{m_{vis}} = 7\fm4 \pm 0\fm5$. P.M. Kilmartin
reported a possible precursor of the nova on a red DSS image at mag
$\mathrm{R} = 17\fm9\pm0\fm4$ and position end figures $02\fs58$,
$55\farcs3$ \citep[in ][]{iauc8123}. Assuming this value, the amplitude
of the outburst reached 10 mags. We have derived the parameters $t_2$
and $t_3$ using a smooth curve fitted to the light curve excluding
those parts which were affected by the minor outbursts. Of the
resulting parameters, $t_2 = 8.6 \pm 1.0$ days,  $t_3 = 32.8 \pm 1.0$
days (formal errors), $t_2$ seems to be more  sensitive to the exact
location of secondary maxima. The light curve is quite fuzzy in the
first $\sim$30 days,  thus the given parameters probably have much
larger uncertainty than what is implied by the formal errors of the 
polynomial fits. Nevertheless, both decline rates indicate that
V4745~Sgr was a  moderately fast nova.

The visual absolute magnitude in maximum was estimated with three 
maximum magnitude versus rate of decline (MMRD) relations using $t_2$ 
(\citealt{dellavalle95,capaccioli89,cohen85}) and two other relations
utilizing $t_3$ \citep{schmidt57,dv78}.  The given $t_2$ resulted in
$-$8\fm75, $-$8\fm79 and $-$8\fm44 for $\mathrm{M}_{\mathrm{V}}$,
respectively, while $t_3$ yielded two other values close to $-$7\fm7.
The $\sim$1 mag difference clearly shows the limitations of the MMRD
relations in complex cases such as the present one. Without any other 
meaningful constraint on the luminosity, we adopt the formal average of
these values, which is $-8\fm3\pm0\fm5$. This leads to an uncorrected
distance modulus $(\mathrm{m} - \mathrm{M})= 15\fm7\pm0\fm7$. 

\subsection{Interstellar reddening and distance}

Despite the large uncertainty of the MMRD-based absolute magnitude,
determining  the interstellar reddening is crucial when calculating
distance limits. Here we  present various empirical pieces of evidence
for a very low reddening toward  V4745~Sgr.

Broad band colour measurements by Gilmore and Kilmartin 
\citep[in][]{iauc8123,iauc8127g,iauc8132g,iauc8160g} gave
$\mathrm{B-V}=-0\fm14 \pm 0\fm03$ at $\Delta t=9~\mathrm{d}$, $-0\fm09$
at $\Delta t=10~\mathrm{d}$, $-0\fm03$ at $\Delta t=16~\mathrm{d}$ and
$+0\fm12$ at $\Delta t=68~\mathrm{d}$. The $(\mathrm{B-V})$ colour of
novae tends to be $\mathrm{B-V}=0\fm23 \pm 0\fm06$ around maximum with
a dispersion of 0\fm16, but two magnitudes below the maximum
the colour dispersion decreases, therefore the relation
$\mathrm{(B-V)_0^{V(max)+2} \approx 0\fm0}$ can be used
\citep{warner95}. The comparison of the given values indicates
negligible photometric reddening.

A common spectroscopic method makes use of interstellar absorption
lines, which do correlate with the reddening \citep[see, e.g.,][ and
references therein]{ciaql}. Unfortunately, the highly sensitive
\citep{munari97} Na~I~D doublet was observed only in the low-resolution
spectra, which had too low resolution for distinguishing narrow
interstellar components (if there were any). We have checked all
medium-resolution spectra for diffuse interstellar bands (DIBs) from
the list of \cite{DIB}, but found none. Therefore, the lack of 
interstellar lines also suggests small reddening.

Another possibility is given by the comparison of the observed and  
theoretical flux ratios of recombination lines. In our spectra the
observed  ratio for Balmer lines $\mathrm{H} \alpha / \mathrm{H} \beta
= 4.4$, which is  not far from the dereddened ratios for V443~Sct 
\citep[see Fig.~6 in ][]{anupama92}. Although this is only a hint for
low reddening, it is consistent with the previous results.

By chance there are two globular clusters, M~70 and NGC~6652, within 
$\sim$1--1.5 degrees of the nova. Both clusters have very low reddening
supporting our findings. For NGC~6652, located $1\degr$ NW of
V4745~Sgr, $\mathrm{E(B-V)}=0\fm09$, while for M~70, located 1\fdg3 NE
of the star,  $\mathrm{E(B-V)}=0\fm07$ was determined \citep{dutra00}.
It appears that despite its low galactic latitude ($\approx-12^\circ$),
V4745~Sgr is located in a low-reddening zone of the Milky Way. The
galactic reddening map by \cite{schlegel98} infers an upper limit of
$\mathrm{E(B-V)}=0\fm118$ for the  position of the nova.

In summary, if there is any reddening toward V4745~Sgr, it must be in
the order of a few hundredths of a magnitude. Consequently, the
distance modulus places the nova between 10 to 19 kpc (assuming zero
reddening) or 9 to 17 kpc (assuming  $\mathrm{E(B-V)}=0\fm1$) from the
Sun. However, the upper limit would imply 3.5 kpc distance from the
galactic plane, which is too large for a classical nova. For that
reason we prefer smaller distances that would place the nova in the
galactic Bulge, just like the very similar transient nova V443~Sct
\citep{anupama92}.

\subsection{The light curve}

\begin{figure}
\centering
\includegraphics[width=88mm]{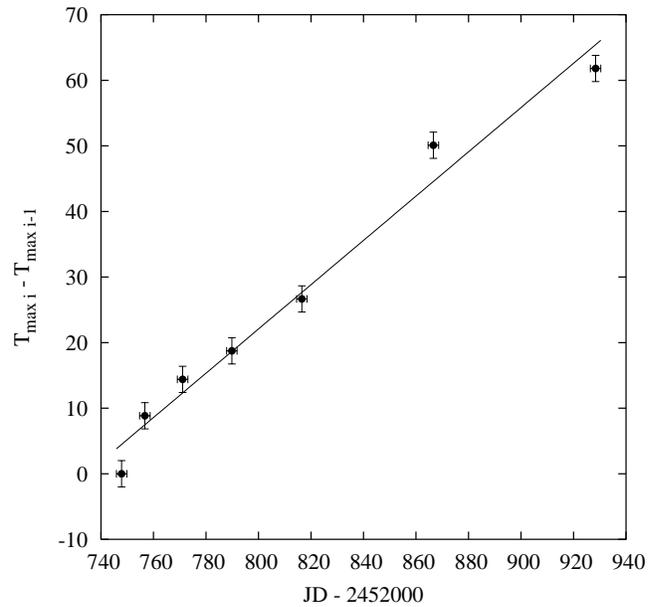}
\caption{Time intervals between the maximum brightness of subsequent
minor outbursts.}
\label{fig:tmax}
\end{figure}

Further interesting details are provided by the light curve. At first
glance, it is apparent that the rebrightenings follow a systematic
pattern in the sense that the recurrence time is steadily increasing.
To characterize it quantitatively, we have determined all epochs of
maxima by fitting low-order polynomials to the selected parts of the
light curve and plotted the consecutive recurrence times versus time. 
As can be seen in Fig.~\ref{fig:tmax}, the time interval between
successive maxima increases monotonically in time. The diagram shows a
quasi-linear trend that is very similar to those of the transient novae
GK~Per and DK~Lac in Figs.\ 1-2 in \cite{bianchini92}. Whatever the
explanation for this  trend is, it shall be accountable for the gradual
increase of the recurrence time. 

\section{Spectroscopic variations and the nature of rebrightenings}

\subsection{A comparison with V1494~Aql}\label{sec:v1494}

A comparable spectroscopic coverage of the transition phase was
presented by \cite{iijima03} for the fast nova V1494~Aql. Its
transition stage started a month after the maximum brightness
\citep{v1494aql}, and the mini-outburst amplitudes were only $\sim
1\fm0 - 1\fm2$ compared to the $\sim 2\fm0$ amplitudes of V4745~Sgr
(Fig.~\ref{fig:lc_spec}). \cite{iijima03} reported that high velocity
broad emission components appeared in the blue and red sides of
H$\alpha$ and H$\beta$ lines near the light maximum of a minor
outburst. They identified these emissions as the trace of high velocity
jets. Similar high velocity jets at H~I and He~II have been also 
observed in supersoft X-ray transients (SXTs)
\citep{becker98,crampton96}. We were unable to identify similar,
symmetric emission features in the emission lines of V4745~Sgr, because
all of these regions are blended with emissions of other elements (see
in Fig.~\ref{fig:jetfig}). 

The main difference between the spectra of V1494~Aql and V4745~Sgr is
the absence of P-Cyg profiles in V1494~Aql during minor outbursts. It
is possible that this difference is due to the inclination difference
between the two systems. V1494~Aql is an eclipsing binary  which shows
fadings with deep minima \citep[][ and references therein]{kato1494},
so that the inclination of the binary system should be close to
90$^\circ$.  In this case, the jets are directed approximately
perpendicularly to the  line of sight (if the jets are perpendicular to
the orbital plane, which, however, can be a very crude approximation),
and we cannot observe the absorbing ejecta  (which forms P-Cyg
profiles) due to the geometry of the system.  On the other hand, for
wide range of the unknown inclination of V4745~Sgr, the hypothetical
jet may be directed toward us, so we can observe P-Cyg profiles on the
blue wings of emission lines, and emissions on the red sides
of lines -- as we can see in Fig~\ref{fig:specevol}b.

X-ray observations of \cite{drake03} also  point towards a possible
connection between the transition phase, jets and the SXTs. 
The detection of supersoft X-rays for V1494~Aql in August 2000, just
after  the transition phase, suggests that hydrogen shell burning was
still  going on the surface of the white dwarf. This fact indicates
that  hydrogen shell burning took place before, during and after the
transition  phase, so that it looks plausible to connect different but
simultaneous  phenomena associated with the nova outburst. However, not
every SXT-like nova shows transition phase. One such example is
V4743~Sgr,  which evolved into the SXT phase in March 2003
\citep{ness03}, six months after its outburst, but its visual
lightcurve beared no transient  phenomena. Unfortunately, we are not
aware of any published X-ray observations of V4745~Sgr and further
comparison based on X-ray data is not possible -- but highly desirable.

\subsection{A comparison with previous models}

How well do our observations fit the various theoretical models
developed for explaining the transition phase?  \cite{retter02} listed
five models offered to the quasi-periodic oscillations during the
transition phase: {\it i)} oscillations of the common envelope that
surrounds the binary system; {\it ii)} dwarf-nova outbursts; {\it iii)}
formation of dust blobs that move in and out of line of sight to the
nova; {\it iv}) oscillations in the wind; {\it v)} stellar oscillations
of the hot white dwarf. \cite{retter02} invoked the effects of the
accretion disc recovery in intermediate polars as an alternative
explanation for the oscillations in the transition phase. As was argued
by \cite{retter02}, the explanation by oscillations of the common
envelope can be easily rejected as that phase lasts less than 1-2 days.
Dwarf nova outbursts can be also ruled out because the accretion discs
in very young post-novae are thermally stable
\citep{retter00,schreiber00}. 

\cite{shaviv01} presented a theory of steady-state super-Eddington
winds, which explains the transition phase by the unstable behaviour of
the wind in a certain stage after the outburst. However, it is apparent
from Figs.\ \ref{fig:specevol}c and \ref{fig:wise_lines} that the
spectrum of V4745~Sgr switched back and forth between being early-type
nova spectrum and nebular one. A few spectral features (e.g. [OIII]
$\lambda\lambda$ 4959,5007, the broad `4640 emission' blend on the one
hand and the Balmer lines, on the other hand) behave very similarly to
those of Nova LMC 1998 No. 2 \citep{sekiguchi89}, with one important
difference: in V4745~Sgr we observed strong P-Cyg profiles with
principal-like absorption systems after some period of time (in the
mezzanine between the minor outbursts),  after the nebular phase has
already developed! In our opinion, these switches contradict the
assumption  of a {\it steady-state} wind, and consequently, the
applicability of the  model by \cite{shaviv01}. 

\cite{retter02} pointed out that several transient novae have been
classified as intermediate polars (IPs) (i.e. the white dwarf has a
moderate magnetic field and thus spins around its axis with a period
shorter than the orbital period) and suggested that there might be a
link between transient novae and IPs.  He proposed that the accretion
discs in IPs are fully destroyed in the nova outburst (contrary to
other systems, for which observational pieces of evidence for early 
presence of the disc were found -- e.g. \citealt{leibowitz92},
\citealt{retter97,retter98}). The re-establishment of the disc and the
subsequent interaction with the magnetosphere of the primary white
dwarf can be violent process that may form strong winds in the inner
part of the disc. \cite{retter02} suggested the oscillations in
transition phase  could be due to a beating of the rotation of the new
accretion disc and the white dwarf magnetic field. The spectroscopic
changes in V4745~Sgr support the presence of a violent process, but
exclude the possibility of rebrightenings caused by a purely
photometric effect. That is why eclipses by dust blobs can also
be safely excluded. 

\cite{schenker} performed a pulsation instability analysis of
envelopes of classical novae and found unstable region that corresponds
to the decay phase of novae. This instability is called
``strange-mode'' (or simply s-mode). \cite{schenker} concluded that
static envelope structures with steadily burning white dwarfs are
highly unstable to radial pulsations over a wide range of physical
parameters. Although the calculated (linear) periods for s-modes are in
the order of hours rather than tens of days, the quasi-static treatment
in \cite{schenker} limits the direct application to the early stages,
which lie in highly non-linear regimes. However, the fast growth rates
and emerging photospheric motion in his calculations indicated that
such instabilities are likely leading to mass loss (Schenker 2004,
personal communication).

Very recently, \cite{cassatella04} presented a spectroscopic
study of the expanding envelope of V1974~Cyg (a regular fast nova with
no transition phase)  based on IUE high-resolution spectra.
Interestingly, they found evidence for two distinct shell ejections, of
which the second one was ejected with a higher velocity, containing
significantly less material. Cassatella et al. suggested that the
second ejection was a reaction to the first one as follows. After the
ejection of the main shell, the envelope is out of balance and will try
to restore equilibrium, but if this does not happen immediately (e.g.
due to some kind of overshooting), a second, less massive shell might
be ejected. Cassatella et al. also pointed out the similarity to the
large eruptions of luminous blue variables, where the main eruption is
followed by a second one \citep{humphreys99}. We might also suggest
that similar repetitive processes may have happened in V4745~Sgr, too.
The contraction and expansion  of the white dwarf envelope would follow
the Kelvin-Helmholtz timescale, which can range from several to several
tens of days, depending on the thermal energy of the envelope and the
luminosity of the given system \cite{haka1}, being roughly in the same
order of magnitude as the timescale of the rebrightenings in
V4745~Sgr.  Since neither the envelope energy nor the luminosity is
known accurately,  at this stage we can only point out the possible
similarity with the hypothetic scenario in \cite{cassatella04}.

\section{Conclusions}

The main conclusions of this paper can be summarized as follows:
\begin{itemize} 

\item We obtained low- and medium-resolution optical spectra  between
$\Delta$t=+26 and +296~d, which revealed remarkable spectroscopic
changes associated with the secondary rebrightenings. During the
optical maxima, the spectrum resembled that of a nova in a very early
stage with broad P-Cyg profiles and principal-like absorptions. Between
maxima, nebular emissions developed and the spectrum was more evolved.
The expansion velocities inferred from the P-Cyg profiles during the
minor outbursts follow a monotonic increase.

\item We estimated the absolute magnitude in maximum with various MMRD
relations. Since the rebrightenings distorted the smooth decline,  the
decline rates are not well defined, so that the applicability of the
MMRD relations is quite limited. 

\item All spectrophotometric reddening indicators (broadband colours, 
interstellar absorption lines, DIBs, emission line flux ratios) 
suggest the interstellar reddening to be negligible ($<0\fm1$ mag). The
nova lies approximately between 9 and 19 kpc, preferably closer to the
lower limit (possibly in the Bulge). 

\item We compared our observational pieces of evidence with
those of another  recent transient nova, V1494~Aql and various models
of the transition phase that can be found  in the literature. We did
not detect jet emissions like those found in V1494~Aql, which may be
due to the different geometry of the system. Of the available models,
radial (s-mode) oscillations in the white dwarf envelope seem to be the
best explanation, although existing calculations cannot be applied
directly to the present case.

\end{itemize}


\begin{acknowledgements} 

This research was supported by the Hungarian OTKA Grants \#T034615 and 
\#T042509; the FKFP Grant 0010/2001 and the Australian
Research Council. We are grateful to the TACs of the Mount Stromlo and
Siding Spring Observatories for allocating telescope time used for this
study. We are also grateful to  Peter Wood for one extra night on the
2.3 m telescope at SSO, which happened to coincide  with the last
observed minor outburst. We would like to express our thanks to the
referee, Prof. Izumi Hachisu,  for his helpful comments.. The NASA ADS
Abstract Service and the SIMBAD Astronomical Database was used to
access data and references.  The computer service of the VSNET group is
also acknowledged.

\end{acknowledgements}

\bibliographystyle{aa}
\bibliography{v4745sgr}

\end{document}